\documentstyle[12pt]{article}
\oddsidemargin--1mm
\begin{document}
\newcommand{\pl}{\partial}
\newcommand{\be}{\begin{equation}}
\newcommand{\ee}{\end{equation}}
\newcommand{\ba}{\begin{eqnarray}}
\newcommand{\ea}{\end{eqnarray}}
\newcommand{\mbf}[1]{\mbox{\boldmath$ #1$}}
\renewcommand{\theequation}{\thesubsubsection.\arabic{equation}}
\renewcommand{\thesubsubsection}{\arabic{subsubsection}}
\def\<{\langle}
\def\>{\rangle}

\begin{center}
{\Large An effective action for monopoles and knot solitons 

\vskip 0.2cm
 in Yang-Mills theory}

\vskip 0.5cm
Sergei V. Shabanov {\footnote{on leave from Laboratory of
Theoretical Physics, JINR, Dubna, Russia}}

\vskip 0.2cm

{\em Departments of Physics and Mathematics, University of Florida,\\
Gainesville, FL- 32611, USA}
\end{center}

\begin{abstract}
By comparision with numerical results in the maximal Abelian projection
of lattice Yang-Mills theory,
it is argued that the nonperturbative dynamics of Yang Mills theory
can be described by a set of fields that take their values 
in the coset space SU(2)/U(1). The Yang-Mills
connection is parameterized in a special way to separate the dependence 
on the coset field. The coset field is then regarded as 
a collective variable, and a method to obtain its effective
action is developed. It is argued that the physical excitations
of the effective action may be knot solitons. A procedure to
calculate the mass scale of knot solitons is discussed for lattice
gauge theories in the maximal Abelian projection. 
The approach is extended to the SU(N) Yang-Mills theory. A relation
between the large N limit and the monopole dominance is pointed out.

\end{abstract}

\subsubsection{Knot solitons and SU(2) Yang-Mills theory}

The action \cite{faddeev1}
\ba
S &=& \int d^4 x\left\{ m^2 (\pl_\mu {\bf n})^2 + H_{\mu\nu}^2\right\}\ ,
\label{1a}\\
H_{\mu\nu}&=&g^{-1}{\bf n}\cdot
(\pl_\mu{\bf n}\times \pl_\nu{\bf n})\ ,
\label{1b}
\ea
where the field ${\bf n}(x)$ is a unit three-dimensional vector, 
${\bf n}\cdot {\bf n} =1$, $m$ is a mass scale
and $g$ is a coupling constant, describes knot solitons in four-dimensional
spacetime. The stability
of solitons is due to the conservative charge known as the Hopf invariant
\cite{faddeev1}.
The knot solitons have a finite
energy and, therefore, may be identified with particle-like
excitations, provided a physical interpretation is given to the ${\bf n}$
field. An interesting relation between the action (\ref{1a}) and the SU(2)
Yang-Mills theory action emerges if one takes the connection of the form
\cite{cho} 
\be
{\bf A}_\mu = g^{-1} \pl_\mu {\bf n}\times {\bf n}\equiv 
{\bf A}_\mu ({\bf n})\ ,
\label{2}    
\ee
where boldface letters
are used to denote isovectors being elements of the adjoint representation
of SU(2), and calculates the corresponding field strength
\be
{\bf F}_{\mu\nu} =\pl_\mu {\bf A}_\nu -\pl_\nu{\bf A}_\mu + g
{\bf A}_\mu\times {\bf A}_\nu\ ,
\label{3}
\ee
one finds that
\be
{\bf F}_{\mu\nu} = {\bf n} H_{\mu\nu}\ .
\label{4}
\ee
That is, the second term of the action (\ref{1a}) is the Yang-Mills
action for the connection of the special form (\ref{2}). It is therefore
rather natural to conjecture that the first term can be generated
by an interaction of the {\em collective} variable ${\bf n}$ with the
other modes of the full Yang-Mills theory \cite{faddeev2,f}. If such 
a conjecture is true, it would mean that the SU(2) quantum  Yang-Mills
theory has particle-like excitations being knot solitons. These excitations
might be good candidates for glueballs.
Since the position of a knot soliton is specified by a contour
in space \cite{fn2,knot}, an effective action 
for the field ${\bf n}$ would also provide
a quantum field description of Polyakov's strings. 

In this letter we discuss a physical interpretation of the field 
${\bf n}$. By analyzing recent developments in lattice gauge
theories, we argue that  the special Yang-Mills connections (\ref{2})
describe the most relevant physical degrees of freedom  of the 
Yang-Mills theory in the confinement phase. We
develop a procedure to calculate their effective action.
The mass scale $m^2$ is described in terms of expectation 
values of some functionals of ${\bf A}_\mu$. We also propose
a numerical procedure to calculate an effective action of
the field ${\bf n}$ using  the Wilson
ensemble of the lattice gauge theory 
in the maximal Abelian projection. The approach is
extended to the SU(N) Yang-Mills theory, where an interesting
relation between dynamics of the coset field and the large
$N$ limit is observed.
 
\subsubsection{A general parameterization of the SU(2) connection}
\setcounter{equation}0

Consider a partition function of the SU(2) Yang-Mills theory
\be
{\cal Z} \sim \int D{\bf A}_\mu\, e^{-S({\bf A})}  \ .
\label{6}
\ee
Here $S({\bf A})$ is the Yang-Mills action.
We assume that some gauge fixing has been made to remove
the divergence of the integral (\ref{6}) caused by the gauge
invariance of the action $S$. For what follows the 
gauge choice is not important.
In the integral (\ref{6}) we want to make a {\em change} of integration
variables
\be
{\bf A}_\mu = g^{-1}\pl_\mu {\bf n}\times {\bf n} + C_\mu {\bf n} +
{\bf W}_\mu\ ,
\label{7}
\ee
where the first two terms is the connection introduced by Cho \cite{cho},
which we denote by $\mbf{\alpha}_\mu$, i.e., ${\bf A}_\mu =
\mbf{\alpha}_\mu + {\bf W}_\mu$, and the isovector ${\bf W}_\mu$ is
perpendicular to ${\bf n}$, that is, ${\bf n}\cdot {\bf W}_\mu =0$.
The idea is then to integrate out $C_\mu$ and ${\bf W}_\mu$ and obtain
an effective action for ${\bf n}$. However, we observe that 
the number of independent field variables in the left hand side of
Eq.(\ref{7}) is 12, while in the right hand side is 14 (4 in $C_\mu$,
2 in ${\bf n}$ and 8 in ${\bf W}_\mu$). To make a change of variables,
we have to impose two more conditions on ${\bf W}_\mu$.

Before doing so, let us analyze the gauge transformation law in the
new variables. A gauge transformation of ${\bf A}_\mu$ contains 
three functional parameters 
\be
\delta {\bf A}_\mu =g^{-1}\pl_\mu \mbf{\omega} + {\bf A}_\mu \times \mbf{\omega}
\equiv g^{-1} \nabla_\mu({\bf A}) \mbf{\omega}\ .   
\label{8}
\ee
The variations $\delta C_\mu$, $\delta{\bf n}$ and $\delta{\bf W}_\mu$
that induce the gauge transformations (\ref{8}) of the 
connection (\ref{7}) should 
depend on five functional parameters because they may also involve variations
under which ${\bf A}_\mu$ does not change at all. Note that
the number of the new variables exceeds that of the old variables 
exactly by 2. Consider a special subset of these
five-parametric transformations which has the form
\be
\delta {\bf n} = {\bf n}\times \mbf{\omega}\ ,\ \ \ 
\delta C_\mu = g^{-1}{\bf n}\cdot \pl_\mu \mbf{\omega}\ ,\ \ \ 
\delta{\bf W}_\mu = {\bf W}_\mu \times \mbf{\omega}\ .
\label{9}
\ee
We have $\delta\mbf{\alpha}_\mu =g^{-1}\nabla_\mu(\mbf{\alpha})\mbf{\omega}$
and, therefore,
the transformations (\ref{9}) induce the gauge transformations (\ref{8}).
The condition ${\bf n}\cdot{\bf W}_\mu =0$ is invariant under the 
transformations (\ref{9}). Hence, the gauge transformed configurations
have the same form (\ref{7}).  

If we impose two additional conditions on ${\bf W}_\mu$ which are
{\em covariant} under the transformations (\ref{9}), then the
transformations (\ref{9}) can be uniquely identified as 
the gauge transformations of the new variables. We propose the 
following conditions
\be 
\nabla_\mu(\mbf{\alpha}){\bf W}_\mu =
\pl_\mu{\bf W}_\mu + gC_\mu{\bf n}\times {\bf W}_\mu +
{\bf n} (\pl_\mu{\bf n}\cdot{\bf W}_\mu) =0\ .
\label{10}
\ee
It is not hard to see that the covariant derivative 
$\nabla_\mu(\mbf{\alpha}){\bf W}_\mu$ transforms as 
an isovector under (\ref{9}). Moreover, taking the dot
product of the right hand side of Eq. (\ref{10}) and
${\bf n}$, we find
${\bf n}\cdot\pl_\mu{\bf W}_\mu +\pl_\mu{\bf n}\cdot{\bf W}_\mu=
\pl_\mu({\bf n}\cdot{\bf W}_\mu)\equiv 0$ because the isovectors
${\bf n}$ and ${\bf W}_\mu$ are perpendicular. Thus, the
right hand side of Eq. (\ref{10}) is an isovector perpendicular to
${\bf n}$ and, therefore, 
the condition (\ref{10}) implies only two independent
conditions on ${\bf W}_\mu$ as required. 

So, we have obtained a change of variables and identified
the gauge transformation law of the new variables. In principle
one can inverse it and find the new variables as functional
of ${\bf A}_\mu$. We will discuss this later upon constructing
the path integral measure for the new variables. One should
point out that there are infinitely many ways to parameterize
the Yang-Mills connection. Natural questions arise. Why
is the parameterization we have chosen so special? Why is
an effective action for the collective variable ${\bf n}=
{\bf n}({\bf A})$ has something to do with the dynamics
of the Yang-Mills theory in the confinement  phase?
Let us discuss these important questions before we turn
to constructing an effective action  for ${\bf n}$. 

\subsubsection{Properties of the new parameterization of the 
SU(2) connection}
\setcounter{equation}0

Eqs. (\ref{7}) and (\ref{10}) determine a complete parameterization 
of a generic SU(2) connection. The parameterization has several
remarkable properties which we are going to discuss.
Consider an isovector ${\bf b}_\mu ={\bf b}_\mu({\bf n})$ which
is constructed of the field ${\bf n}$ and its derivatives so that
${\bf b}_\mu\cdot{\bf n}=0$. Since $\pl_\mu{\bf n}\cdot{\bf n}=0$,
such an isovector can always be taken as a linear combinations
${\bf b}_\mu =b_{\mu\nu}\pl_\nu{\bf n}$ with the coefficients
$b_{\mu\nu}$ being functions of ${\bf n}$ and its derivatives.
Let us construct a complex scalar field
\be
\Phi = {\bf b}_\mu\cdot {\bf W}_\mu + 
i{\bf n}\cdot({\bf b}_\mu\times {\bf W}_\mu)\ .
\label{11}
\ee
Consider gauge transformations that leave the field ${\bf n}$
unchanged. These are rotations about ${\bf n}$, i.e., 
$\mbf{\omega} = \xi{\bf n}$ in (\ref{9}). We get $\delta C_\mu
=g^{-1}\pl_\mu\xi$ and $\delta\Phi = i\xi\Phi$. That is, 
the field $C_\mu$ is a Maxwell field with respect to the gauge
subgroup U(1) which is a stationary group of ${\bf n}$,
while the scalar field $\Phi$ plays the role of a charged 
field. If for some reasons we would like to describe the SU(2)
Yang-Mills theory as an effective QED, the set of fields 
$C_\mu$, $\Phi$ and ${\bf n}$ is just enough to carry {\em all} 
physical degrees of freedom of the original SU(2) theory.
This can be understood as follows. The field ${\bf W}_\mu$ 
has six independent components in our parameterization of the 
SU(2) connection. Using the isotopic rotations we may impose
two gauge conditions on ${\bf W}_\mu$ 
to break SU(2) to U(1). 
Among the remaining four components of ${\bf W}_\mu$
we can always select two (real) components which are to be identified 
with the complex scalar field $\Phi$ by a suitable choice of ${\bf b}_\mu$. 
The other two components are uniquely fixed by the Gauss law
$\delta S/\delta{\bf A}_0 =0$. Recall that
the SU(2) gauge theory has three Lagrange multipliers ${\bf A}_0$.
If we impose a gauge condition on the dynamical variables ${\bf A}_i$
$(i=1,2,3)$ with the aim to get a dynamical description only in terms
of the physical degrees of freedom, then the Lagrange multipliers
must be obtained by {\em solving} 
the Gauss law for ${\bf A}_0$ in the gauge (or {\em parameterization}) 
chosen for ${\bf A}_i$. The Gauss law imposes a restriction
on possible parameterizations of connections ${\bf A}_\mu$ via 
only physical variables.  
Since U(1) is left unbroken as the gauge group of the theory, only two
equations in the Gauss law are to be solved so that the Lagrange
multiplier $C_0$ associated with the U(1) symmetry  does not get fixed.
Thus, in such an effective Abelian gauge theory, which is dynamically 
equivalent to the non-Abelian SU(2) theory, the field ${\bf W}_\mu$
carries only two physical degrees of freedom associated with
the complex scalar field (\ref{11}). Here only an effective dynamics
of ${\bf n}$ (an effective field theory for Polyakov's strings) 
will be discussed and, therefore, an explicit parameterization
of ${\bf W}_\mu$ via physical variables is not relevant. 
Such a parameterization is 
important to formulate duality properties 
in Yang-Mills theory \cite{faddeev2,f}.

What is the physical meaning of the field ${\bf n}$ in the effective
Abelian gauge theory described above? To answer this question let us make an 
Abelian projection (in 't Hooft's terminology \cite{thooft}) 
of the Yang-Mills theory by imposing
a gauge on the field ${\bf n}$ rather than on ${\bf W}_\mu$. 
By a suitable gauge rotation of the
isovector ${\bf n}$ we can always direct it along, say, the third
coordinate axis in the isospace, i.e., ${\bf n}={\bf n}_0=(0,0,1)$.
In this case the field ${\bf W}_\mu$ would have four physical
degrees of freedom (in the sense of the Hamiltonian formalism
as described above). Moreover, since
the conditions (\ref{10}) are covariant under gauge transformations,
the gauge transformed fields $C_\mu$, 
${\bf W}_\mu$ and ${\bf n}$ should also satisfy them. Setting
${\bf n}$ equal to ${\bf n}_0$ in (\ref{10}), the latter turns
into the condition which is well known in lattice Yang-Mills theories 
as the maximal Abelian gauge \cite{mag}. 

The reason of why the maximal Abelian gauge 
is so special in lattice Yang-Mills theory is the following. Suppose
configurations ${\bf A}_\mu^{q}$, $q=1,2,...,Q$, are elements 
of the Wilson ensemble, i.e., they are generated by means of the 
Monte-Carlo method with the 
Boltzmann probability $\exp[-S({\bf A})]$. An expectation
value of any quantity $F({\bf A})$ is 
$\<F\>=Q^{-1}\sum_qF({\bf A}^q)$. For example, one can take
$F=W_C$ to be the Wilson loop. From $\<W_C\>$ one finds the string
tension $\sigma_{su(2)}$ which is a coefficient 
in the linearly rising part of the potential between a
heavy quark and antiquark. 
For every configuration ${\bf A}^q_\mu$
one can find a gauge transformation $U=U({\bf A}^q)$ such that
the gauge transformed configuration has the form ${}^{U}\!{\bf A}_\mu^q=
{\bf n}_0{}^U\! C_\mu^q + {}^U{\bf W}_\mu^q$ where 
${}^U{\bf W}_\mu^q$ and ${}^U\! C_\mu^q$ 
satisfy the condition (\ref{10}) with ${\bf n}={\bf n}_0$. That is, every 
configuration from the Wilson ensemble have been gauge transformed to
satisfy the maximal Abelian gauge. Now one takes only the Abelian 
parts ${\bf n}_0{}^U\! C_\mu^q$ of ${}^{U}\!{\bf A}_\mu^q$ and use them
to calculate $\<W_C\>$ again. As ${}^U\! C_\mu^q$ are Abelian configurations
(the action is quadratic in $C_\mu$), one would expect 
the perimeter law corresponding to the Coulomb interaction of static sources
rather than the area law, i.e., one would expect 
the corresponding string tension $\sigma_{u(1)}$ to vanish. 
A surprising numerical result is that such a procedure gives
the area law again, and $\sigma_{u(1)}$ is nearly the same
as the full string tension $\sigma_{su(2)}$ (to be exact,
it is 92 per cent of $\sigma_{su(2)}$).
The phenomenon is called the Abelian dominance \cite{ad,cam}. It might look
rather mysterious especially in view of that $W_C({\bf A})$ is 
a {\em gauge invariant} quantity.

The paradox disappears if one observes that
the gauge transformations $U({\bf A}^q)$ that are used to 
implement the maximal Abelian gauge are {\em not regular} in spacetime
(e.g., $U^\dagger[\pl_\mu,\pl_\nu]U\neq 0$). Strictly speaking,
they are {\em not} gauge transformations because the tensor
quantities, like, e.g., the field strength, are no longer
transformed homogeneously. For this reason
the Abelian vector potential ${}^U\! C_\mu^q =
{\bf n}_0\cdot{}^{U}\!{\bf A}_\mu^q$
has singularities (or defects) which have quantum numbers of Dirac
magnetic monopoles with respect to the unbroken U(1) gauge group.
Therefore the Maxwell vector potential ${}^U\! C_\mu^q$  
describes dynamics of photons {\em and} also that of Dirac magnetic monopoles
which are physical degrees of freedom of the Wilson ensemble
in the maximal Abelian gauge. 
\footnote{In fact, this is the case for any Abelian projection \cite{thooft},
but not for every Abelian projection the Abelian 
(or monopole) dominance holds \cite{cam}.}

Thus, the {\em singular} transformations $U({\bf A}^q)$ transfer
some relevant {\em physical} degrees of freedom from the 
non-Abelian components
(perpendicular to ${\bf n}_0$) of the connection
to the Abelian ones (parallel to ${\bf n}_0$).
These degrees of freedom have been identified as Dirac magnetic monopoles.
Locations of monopoles 
can be found by studying the magnetic flux carried by ${}^U\! C_\mu^q$ through
surfaces around each dual lattice cite \cite{mag}.
Since regular (photon) configurations of ${}^U\! C_\mu^q$ cannot provide
the area law for the Wilson loop, the monopole part of ${}^U\! C_\mu^q$ must 
be expected to give a major contribution to $\sigma_{u(1)}\approx
\sigma_{su(2)}$. If one removes the photon part from ${}^U\! C_\mu^q$
and calculate $\<W_C\>$ using only the monopole part of ${}^U\! C_\mu^q$,
the corresponding string tension $\sigma_m$ differs from $\sigma_{u(1)}$
only by 5 per cent, i.e., $\sigma_m\approx\sigma_{u(1)}\approx\sigma_{su(2)}$.
This is known as the monopole dominance \cite{md,cam}. 

The conclusion one obviously arrives at is that the nonperturbative
Yang-Mills dynamics favors configurations which look like Dirac
magnetic monopoles in the maximal Abelian gauge. 
An interpretation of these degrees of freedom as Dirac monopoles
is gauge dependent, but the very fact of their existence
is certainly gauge independent.
These configurations capture the 
most relevant degrees of freedom of the Yang-Mills theory
in the confinement phase.
This is a nontrivial dynamical statement 
discovered  in lattice gauge theories.
Note that an explicit form of ${\bf A}_\mu^q$ is determined by the 
{\em full} Yang-Mills action, and it is by no means obvious that
the dynamics must be such that the configurations that saturate
the area law should look like Dirac magnetic monopoles in some 
particular gauge. 

In our parameterization of the SU(2) connection the maximal Abelian
gauge is a simple {\em algebraic} gauge ${\bf n}={\bf n}_0$. Therefore
we expect that for some connections ${\bf A}_\mu^q$
the gauge transformation which transforms ${\bf n}({\bf A}^q)$ to the
special form ${\bf n}_0$ is not regular, and the gauge transformed 
Abelian potential ${}^U\! C_\mu^q$ should have singularities which have
quantum numbers of Dirac monopoles with respect
residual gauge group U(1) (rotations about ${\bf n}_0$). 
As the condition (\ref{10}) is covariant under the gauge
transformations (\ref{9}) and turns into the maximal Abelian
gauge if we gauge transform 
 ${\bf n}\rightarrow {\bf n}_0$, $C_\mu \rightarrow
 {}^U\! C_\mu^q ={\bf n}_0\cdot {}^U\!{\bf A}_\mu^q$
and ${\bf W}_\mu\rightarrow {}^U{\bf W}_\mu^q = {}^U\!{\bf A}_\mu^q -
{\bf n}_0 ({\bf n}_0\cdot{}^U\!{\bf A}_\mu^q)$, where
$U({\bf A}^q)\in SU(2)/U(1)$ is defined so that 
${}^U\!{\bf A}_\mu^q$ satisfies the maximal Abelian gauge,
we conclude that our field ${\bf n}$  
should have the form
\be
{\bf n}({\bf A}^q) = \frac 12{\rm tr}\,\left[
\mbf{\tau}U^\dagger({\bf A}^q)\tau_3U({\bf A}^q)\right]\ .
\label{na}
\ee
Here the components of the isovector $\mbf{\tau}$ are
the Pauli matrices, ${\rm tr}\,(\tau_a\tau_b)=2\delta_{ab}$. 
The relation (\ref{na}) 
determines the configurations of the field ${\bf n}$
for the Wilson ensemble. The ensemble (\ref{na}) contains
all information about dynamics of the monopole
degrees of freedom of the maximal Abelian projection {\em by construction}.

In the continuum theory it is also easy
to give examples of ${\bf n}$ with such properties. If we set
${\bf n}({\bf x},t)=g{\bf x}/r$, where $r=|{\bf x}|$, then the configuration
(\ref{2}) is the famous non-Abelian monopole of Wu and Yang \cite{wu}.
Observe that the Wu-Yang monopole corresponds to zero configurations
of $C_\mu$ and ${\bf W}_\mu$ in our parameterization. 
Now we can implement the maximal Abelian projection.
We can always find an orthogonal matrix that
transforms ${\bf n}$ to the special form ${\bf n}_0=(0,0,1)$.
This gauge transformation creates nonzero $C_\mu$ according
to (\ref{9}). The gauge transformed configuration becomes purely
Abelian, and $C_\mu$ describes a Dirac magnetic monopole localized
at the origin. Note that the orthogonal matrix used to
direct ${\bf n}$ along the third coordinate axis is singular. 
This singularity generates a singularity of $C_\mu$ on the Dirac string
extended along the negative part of the third coordinate axis.
In general, we can set ${\bf n} =\mbf{\phi}/\phi$,
where $\phi$ is the norm of an isovector field $\mbf{\phi}$.
If $\mbf{\phi}(x)=0$, then a solution to this equation is a collection of
worldlines $x_\mu=x_\mu^p(s)$ which are identified with the wordlines 
of Dirac monopoles carried by $C_\mu$ after the singular gauge transformation
${\bf n}\rightarrow{\bf n}_0$:
\be
g^{-1}\pl_\mu{\bf n}\times{\bf n} +C_\mu{\bf n}
\rightarrow \left[g^{-1}{\bf n}_0\cdot(\pl_\mu\mbf{\xi}\times\mbf{\xi})
+C_\mu\right]{\bf n}_0 = {}^U\! C_\mu
{\bf n}_0 \equiv (C_\mu^\xi +C_\mu){\bf n}_0\ ,
\label{ap}
\ee 
where $\mbf{\xi}=(\sin(\theta/2)\cos\varphi,\sin(\theta/2)\sin\varphi,
\cos(\theta/2))$ if ${\bf n}=(\sin\theta\cos\varphi,\sin\theta\sin\varphi,
\cos\theta)$. The first term $C_\mu^\xi$ in ${}^U\! C_\mu$  is the vector
potential of magnetic monopoles in the maximal Abelian projection.

Thus, the collective variable ${\bf n}({\bf A})$ captures the 
degrees of freedom  responsible for the Abelian or monopole
dominance discovered in lattice gauge theories. According
to lattice simulations, in the confinement phase the monopole 
current (in the maximal Abelian projection) is dense, while
in the deconfinement phase it is very dilute \cite{cam}. Frankly speaking,
the monopoles condense in the confinement phase,
and one needs a {\em field} (an order parameter) 
to describe their dynamics. A single Maxwell field ${}^U\! C_\mu$ cannot
describe photons and monopoles simultaneously because it  
has no independent components available as the ``monopole'' field.
Note that ${}^U\! C_\mu$ is a sum of the photon field $C_\mu$ and
the monopole field $C_\mu^\xi$ in the maximal Abelian projection (\ref{ap}).   
Now we recall that the ``monopole'' interpretation of the physical
degrees of freedom of Yang-Mills theory which saturate the area law 
is associated with a specific gauge. But once 
these degrees of freedom have been identified as those
of the field ${\bf n}$, there is no need in that gauge anymore.
Instead of imposing the (maximal Abelian) gauge ${\bf n}={\bf n}_0$
and facing a hard problem to find a quantum field description 
of monopole defects in ${}^U\! C_\mu$, we say that 
in the confinement phase the collective field ${\bf n}={\bf n}({\bf A})$
should be identified as the most relevant degree of freedom.
Therefore its effective action should capture the main features 
of the nonperturbative Yang-Mills dynamics.
This is a gauge {\em invariant}
approach to the ``monopole'' dynamics because it uses only 
a reparameterization of the original Yang-Mills dynamical variables
via new collective variables and no gauge fixing. The effective dynamics
of ${\bf n}$ can be studied in any Abelian projection (gauge) if so
desired.

The action in the variables (\ref{7}) has the form
\be
S({\bf A}) = \frac{1}{4}\left\{G_{\mu\nu}^2 +  
2(\nabla_\mu{\bf W}_\nu)^2 +
4(G_{\mu\nu}{\bf n}
+\nabla_\mu{\bf W}_\nu)\cdot {\bf W}_{\mu\nu} 
+ {\bf W}_{\mu\nu}^2\right\}\ ,
\label{12}
\ee
where $\nabla_\mu=\nabla_\mu(\mbf{\alpha})$, 
${\bf W}_{\mu\nu} ={\bf W}_\mu\times{\bf W}_\nu$ and
$G_{\mu\nu}=\pl_\mu C_\nu-\pl_\nu C_\mu -H_{\mu\nu}$.
It is quadratic in $C_\mu$ so the Maxwell field 
can be integrated out, while the integral over ${\bf W}_\mu$
can only be done perturbatively, or in a stationary phase
approximation by invoking instanton solutions, or numerically. 
As a point of fact, lattice simulations
show a correlation between monopole-antimonopole loops
and instantons \cite{im}. So an instanton induced interaction of
the field ${\bf n}$ might be an important piece of the 
effective action of ${\bf n}$.
In this regard it is noteworthy \cite{f} that
Witten's multi-instanton Ansatz \cite{witten} can be written
in our parameterization of the SU(2) connection as follows
${\bf n}=g{\bf x}/r$, ${\bf W}_0=0$, and 
\be
{\bf W}_i=\varphi_2(r,t)\pl_i{\bf n}\times{\bf n}
+\varphi_1(r,t)\pl_i{\bf n}\ ,\ \  
C_i=B_1(r,t)n_i\ ,\ \ C_0 =B_2(r,t)\ .
\label{13}
\ee
Here $B_{1,2}$ and $\varphi_{1,2}$ are, respectively, two-dimensional
electromagnetic and scalar fields introduced by Witten.
Note that the condition (\ref{10}) is satisfied {\em identically}
for the Ansatz (\ref{13}).

\subsubsection{An effective action for the ${\bf n}$ field}
\setcounter{equation}0

To construct an effective action for the field ${\bf n}$, we observe that
the change of variables proposed in section 2 allows us to find
two equations which determine ${\bf n}$ as an {\em implicit}
functional of ${\bf A}_\mu$. Indeed, we have $C_\mu = {\bf n}\cdot{\bf A}_\mu$.
Therefore 
\be
{\bf W}_\mu = {\bf A}_\mu -{\bf n} ({\bf A}_\mu\cdot{\bf n}) -g^{-1}
\pl_\mu{\bf n}\times {\bf n} \equiv {\bf A}_\mu -
\mbf{\alpha}_\mu({\bf A},{\bf n})\ .
\label{14}
\ee 
The field ${\bf W}_\mu$ should satisfy the condition (\ref{10}), which
leads to the desired equations for ${\bf n}={\bf n}({\bf A})$:
\be
\mbf{\chi}({\mbf A},{\bf n}) \equiv \nabla_\mu({\bf A})\mbf{\alpha}_\mu
({\bf A},{\bf n}) -\pl_\mu{\bf A}_\mu =0\ .
\label{15}
\ee
It is easy to verify that ${\bf n}\cdot\mbf{\chi}\equiv 0$. So the
isovector $\mbf{\chi}$ is always perpendicular to ${\bf n}$ and,
hence, Eq. (\ref{15}) contains only two independent equations for
two independent components of ${\bf n}$. Consider the functional
$\Delta =\Delta({\bf A},{\bf n})$ defined by the equation
\be
1 = \int D{\bf n}\, \Delta ({\bf A},{\bf n})\,
\delta (\mbf{\chi})\ .
\label{16}
\ee
Substituting the identity (\ref{16}) into (\ref{6}) we find
an equivalent representation of the Yang-Mills partition function
\ba
{\cal Z} &\sim& \int D{\bf n}\,e^{-S_{eff}({\bf n})}\ ,\label{17a}\\
S_{eff}({\bf n}) &=& -\ln\int D{\bf A}\,\Delta ({\bf A},{\bf n})\,
\delta (\mbf{\chi})\, e^{-S({\bf A})}\ .\label{17b}
\ea
The integral over Yang-Mills fields in (\ref{17b}) could be done 
analytically only
either by perturbation theory or by means of
instantons. In the latter case the functional integral
is replaced by an ordinary integral over the instanton moduli space,
while ${\bf A}_\mu$ is replaced by an instanton configuration.
To develop perturbation theory, one should write 
the functional $\Delta$ as an integral over ghost fields.
From the definition (\ref{16}) of $\Delta$ it follows that
$\Delta = \det(\delta\mbf{\chi}/\delta{\bf n})$. Introducing 
a set of complex ghost fields $\mbf{\eta}$ such that 
$\mbf{\eta}\cdot{\bf n}=0$, we have
\ba
\Delta({\bf A},{\bf n}) &=& \int D\mbf{\eta}^\dagger D\mbf{\eta}
\, \exp\left\{-S_{gh}(\mbf{\eta}^\dagger,\mbf{\eta},{\bf A}, {\bf n})
\right\}\ ,
\label{18a}\\
S_{gh}&=&\int\! d^4x\, \mbf{\eta}^\dagger \cdot
\nabla_\mu({\bf A})\left[\mbf{\eta} ({\bf n}\cdot{\bf A}_\mu)
+{\bf n}(\mbf{\eta}\cdot{\bf A}_\mu) +\pl_\mu{\bf n}\times\mbf{\eta}
-{\bf n}\times \pl_\mu\mbf{\eta}\right]\ .
\label{18b}
\ea
The delta function of $\mbf{\chi}$ in (\ref{17b}) can be written
in the exponential form via the Fourier transform. 

The ghosts introduced have nothing to do
with a gauge fixing. As has been pointed out before, any gauge fixing 
condition can be assumed in (\ref{6}) by an appropriate 
modification of the measure $D{\bf A}_\mu$. 
Note that the integral (\ref{17b}) is invariant under 
{\em simultaneous} gauge transformations of ${\bf A}_\mu$ and
${\bf n}$ (cf. (\ref{8}) and (\ref{9})). If the ghost representation  
(\ref{18a}), (\ref{18b}) is used, then the ghost field is transformed
as $\delta\mbf{\eta} =\mbf{\eta}\times\mbf{\omega}$.  By imposing
a gauge on ${\bf A}_\mu$, this gauge freedom is removed. Therefore 
a gauge can be chosen to make computation of the integral
(\ref{17b}) convenient (e.g., the Lorentz gauge for
perturbation theory, or a background gauge if the instanton
technique is used). Accordingly, there will also be a conventional
set of ghost fields associated with the gauge chosen.

It is convenient to make a shift of the integration variables
${\bf A}_\mu \rightarrow {\bf A}_\mu +g^{-1}\pl_\mu{\bf n}
\times {\bf n}$ to extract the tree level contribution
to the effective action (the second term in (\ref{1a})).
If the effective action (\ref{17b}) supports knot solitons
as collective excitations, then   
in the gradient expansion of (\ref{17b}) there should exist
the mass scale $m^2\neq 0$ which is determined by the equation
\be 
\left.\frac{\delta^2S_{eff}({\bf n})}{\delta n_a(x)\delta n_b(y)}
\right\vert_{\pl {\bf n}=0} = -m^2\delta_{ab}\pl_\mu^2\delta(x-y)\ .
\label{19}
\ee
Substituting (\ref{17b}) into (\ref{19}) we obtain 
the mass scale in terms of expectation values of certain
operators in the Yang-Mills theory. 
As has been argued in \cite{faddeev2}, the mass scale term 
must be a leading term in the gradient expansion of the effective
action $S_{eff}({\bf n})$. However, it may also appear to be zero.
Since there is no 
monopole dominance in the perturbative regime, there 
must be a nonperturbative input into calculation of $m^2$.

One way to do so is to compute $m^2$ numerically.
Suppose we have  a Wilson ensemble of ${\bf A}_\mu$.
For every configuration ${\bf A}_\mu$ one can calculate the 
group element $U({\bf A})$ such that ${}^U\!{\bf A}_\mu$
satisfies the maximal Abelian gauge. Next, by means of 
(\ref{na}) one can obtain an ensemble of the
field ${\bf n}$. The problem is to find the Boltzmann
probability $\exp(-S_{eff}({\bf n}))$ which generates
the ensemble of ${\bf n}$. Such a problem can be solved
by the so called inverse Monte-Carlo method \cite{imc}.
This method has recently been applied to calculate 
an effective action of the monopole current in
the maximal Abelian projection \cite{imcj}. It would be
interesting to apply this numerical technique to prove
the existence of the mass scale $m^2$. The radiative 
corrections to the action (\ref{1a}) 
can be obtained by perturbation theory proposed above. 

The mass scale term in (\ref{1a}) looks like a mass
term in the  (reduced) Yang-Mills theory of the 
special connection (\ref{2})
\be
m^2\pl_\mu{\bf n}\cdot\pl_\mu{\bf n} = g^2m^2{\bf A}^2_\mu({\bf n})\ .
\label{20}
\ee
This suggests also that the inverse Monte-Carlo method can be
applied to the ensemble of ${\bf A}_\mu({\bf n})$ directly.
If one takes the ensemble of monopole connections ${\bf n}_0C_\mu^\xi$
in the maximal Abelian projection (i.e., after the projection
all off-diagonal components of the connections 
as well as the photon part in the
diagonal components are set to zero), then the ensemble of
${\bf A}_\mu({\bf n})$ can be obtained by gauge transformations
$U^\dagger({\bf A})$ of ${\bf n}_0C_\mu^\xi$ because by construction 
(\ref{ap}) we have ${}^U\!{\bf A}_\mu({\bf n}) = {\bf n}_0C^\xi_\mu$.
Here $U({\bf A})$ are gauge group elements which are used to 
implement the maximal Abelian gauge on the Wilson ensemble.

\subsubsection{A generalization to the SU(N) gauge group}
\setcounter{equation}0

A realistic theory has the gauge group SU(3). The monopole
dominance has also been established for the SU(3)  lattice gauge
theory in the maximal Abelian projection (see, e.g., \cite{cam}
and references therein). Here we construct a parameterization
of the SU(N) connection such that the ``monopole'' degrees of
freedom in the maximal Abelian projection are described 
by a coset field $SU(N)/[U(1)]^{N-1}$.
The parameterization is similar, but not the same, to
the parameterization proposed recently
by Faddeev and Niemi \cite{sun1}, and it also differs from 
the parameterization of Periwal \cite{sun2}.
To develop an effective action of the coset field, we
follow the method discussed above for the SU(2) case, meaning
that no explicit elimination of nonphysical degrees of freedom
is made, and thereby a hard problem (unsolved in \cite{sun1,sun2})
of solving the Gauss law is avoided. We shall point out an interesting
relation emerging between the large N limit and the monopole
dominance.

Let $h_i$, $i=1,2,...,N-1$, be a basis of the Cartan subalgebra
of the algebra su(N). We assume the basis to be orthonormal
with respect to the Killing form $(h_i,h_k)=\delta_{ik}$.
Recall that for any two elements of a Lie algebra
the Killing form is defined as $(y,z) ={\rm tr}(\hat{y}\hat{z})$,
where $\hat{y}z=[y,z]$ and $[,]$ is a Lie product in the Lie
algebra. In a matrix representation $(y,z)= c\,{\rm tr}(yz)$,
where $c$ depends on the Lie algebra \cite{zhel} ($c=2N$
for su(N)). Consider an orthonormal
Cartan-Weyl basis \cite{zhel} 
so that every Lie algebra element can be decomposed as
\be 
\omega = \sum_{\beta>0}\left(\omega_\beta^c c_\beta +\omega_\beta^s
s_\beta\right) + \sum_k \omega_kh_k\ .
\label{c1}
\ee
Here $\beta$ ranges over all positive roots of the algebra, and
the coefficients $\omega_{\beta}^{c,s}, \omega_k$ are real.
In what follows we will only need the commutation relations
$[h,c_\beta]=i(h,\beta)s_\beta$ and $[h,s_\beta]=-i(h,\beta)c_\beta$
(for any element $h$ from the Cartan subalgebra). So the 
other commutation relations of the basis elements are omitted. They can
be found in, e.g., the textbook \cite{zhel}. Let $U(x)\in
SU(N)/[U(1)]^{N-1}$ and $n_k=U^\dagger h_kU$. Since the Killing
form is invariant under the adjoint action of the group,
we conclude $(n_i,n_k)=\delta_{ik}$. We set (cf. \cite{sun1})
\be
A_\mu = ig^{-1}N[\pl_\mu n_k,n_k] + n_kC^k_\mu +W_\mu \equiv
\alpha_\mu + W_\mu\ ,
\label{c2}
\ee
where $(n_k,W_\mu)=0$ and $W_\mu$ also satisfies the following
$N^2-N$ conditions
\be
\nabla_\mu(\alpha)W_\mu \equiv \pl_\mu W_\mu + igC_\mu^k[n_k,W_\mu]
-N\left[[\pl_\mu n_k,n_k],W_\mu\right] =0\ .
\label{c3}
\ee
The connection $A_\mu$ has $4(N^2-1)$ independent components
which are now represented by $N^2-N$ independent functions
in $n_k$, $4(N-1)$ functions in $C_\mu^k$ and $4(N^2-1)-(N^2-N) - 4(N-1)$
functions in $W_\mu$. 

The gauge transformation law in the new variables reads
\be
\delta n_k = i[n_k,\omega]\ ,\ \ \ 
\delta W_\mu = i[W_\mu, \omega]\ ,\ \ \ 
\delta C_\mu^k = g^{-1}(n_k, \pl_\mu\omega)\ .
\label{c4}
\ee
To prove this, we have to show that the transformations (\ref{c4})
induce an infinitesimal gauge transformation of $A_\mu$:
$\delta A_\mu = g^{-1}\nabla_\mu(A)\omega$. Comparing the 
gauge transformations of the left- and right-hand sides of
Eq. (\ref{c2}), we find that the relation 
\be
\pl_\mu \omega = N\left[n_k,[n_k,
\pl_\mu\omega]\right] + n_k (n_k,\pl_\mu\omega)
\label{c5}
\ee
has to hold true for any $\pl_\mu\omega$. 
Consider a {\em local} Cartan-Weyl orthonormal basis 
$c_\beta^U = U^\dagger c_\beta U$, $s_\beta^U = U^\dagger s_\beta U$
and $n_k$. We decompose the element $\pl_\mu\omega$ in this
local basis and compute the commutators in the l.h.s. of Eq. (\ref{c5})
by means of the Cartan-Weyl commutation relations:
$[n_k,[n_k,s_\beta^U]]=-i(h_k,\beta)[n_k,c_\beta^U]=(h_k,\beta)^2 s_\beta^U$
and, similarly, $[n_k,[n_k,c_\beta^U]]=(h_k,\beta)^2 c_\beta^U$, while
$[n_k,n_j]=0$.
Relation (\ref{c5}) follows from the resolution of unity in the
local Cartan-Weyl basis, provided $\sum_k(h_k,\beta)^2= 1/N$
for every root $\beta$. As $h_k$ form an orthonormal basis 
in the Cartan subalgebra with
respect to the Killing form, we have $\sum_k(h_k,\beta)^2=
(\beta,\beta) ={\rm tr}\hat{\beta}^2 =1/N$. The latter means that
all roots of su(N) should have the same norm, which is 
easily established from the Dynkin diagram for su(N).
To compute the norm, we set $(\beta,\beta) = b$ for any root 
and calculate the matrix
elements of $\hat{\beta}$ by applying it to the basis elements
of the Cartan-Weyl basis, then from the relation $
(\beta,\beta) ={\rm tr}\hat{\beta}^2$ and the root pattern it follows
that $b=1/N$. Thus, relation  (\ref{c5}) is a true identity.
In a similar fashion, one can prove that $(n_k,\nabla_\mu(\alpha)W_\mu)
=\pl_\mu(n_k,W_\mu)\equiv 0$. The idea is to compute the
Abelian components of the last term in (\ref{c3}) in the local
Cartan-Weyl basis. The Abelian components can only be produced
by those $c_\beta-$ and $s_\beta-$components of $W_\mu$
and $[\pl_\mu n_k,n_k]$ that correspond to the {\em same}
root because \cite{zhel} $[c_\beta ,s_\beta]=i\beta$, but
$(h,[c_\beta,s_{\beta'}])=(h,[s_\beta,s_{\beta'}])=
(h,[c_\beta,c_{\beta'}])=0$ for any $\beta\neq \beta'$
and any element $h$ of the Cartan subalgebra.
Thus, Eq. (\ref{c3}) indeed contains
only $N^2-1 -(N-1)=N^2-N$ independent conditions on $W_\mu$.

Now it is easy to see that in our parameterization the maximal
Abelian gauge is an {\em algebraic} gauge $n_k(x)=h_k$. Given
a connection $A_\mu$, one can find $U(A) \in SU(N)/[U(1)]^{N-1}$
such that ${}^U\! A_\mu$ satisfies the maximal Abelian gauge.
The Maxwell fields in the maximal Abelian projection ${}^U\!
C_\mu^k = (h_k,{}^U\! A_\mu)$ carry Dirac monopoles. By construction
these ``monopole'' degrees of freedom are described by the
coset fields $n_k(A) =U^\dagger(A)h_kU(A)$ in our parameterization.
If $\phi(x)$ is a local operator that transforms in the adjoint
representation of the gauge group, then by a gauge transformation
it can always be brought to the Cartan subalgebra. Such a gauge
transformation will be singular at spacetime points 
where a special gauge invariant polynomial of $\phi$ 
has zeros \cite{svs}. This polynomial is the Jacobian
of the change of variables $\phi =U^{-1}hU$ where 
$h$ is an element of the Cartan subalgebra. In particular,
for the SU(N) group, the singularities form worldlines
which can be identified with the worldlines of magnetic monopoles 
in the effective $[U(1)]^{N-1}$ Abelian gauge theory \cite{svs}. Therefore,
if $n_k\sim \phi/(\phi,\phi)^{1/2}$, the gauge transformation
$n_k\rightarrow h_k$ is singular, in general, and creates Dirac magnetic
monopoles in the Maxwell fields ${}^U\! C_\mu^k$.

An effective action of the fields $n_k$ can be calculated 
similarly to the case of SU(2) discussed in section 4.
A generalization is straightforward, so we omit the details.
A few concluding remarks are in order. A remarkable fact
that all roots of SU(N) have the same norm was crucial
to establish a relation between the ``monopole'' configurations
of the maximal Abelian projection and the parameterization
(\ref{c2}) of the SU(N) connection. Clearly, for any other
compact gauge group the parameterization (\ref{c2}),
(\ref{c3}) is not applicable. This
should not be regarded as a drawback. Note that only for
SU(N), Abelian projections lead to monopole-like defects
in gauge fields \cite{thooft}. For instance, for SO(N)
the defects would be objects extended in space, like
Nielsen-Olsen strings \cite{thooft}. So it is natural
to expect a different form of the connection that
captures such degrees of freedom. 

It is also rather interesting that the ``monopole''
part of the SU(N) connection (\ref{c2}) has the factor $N$.
So in the large $N$ limit it has to play the role
of ``dominant'' degrees of freedom, which implies that
there exists a relation between the large N limit
and the monopole dominance in gauge theories.
Finally, we remark that an effective action for the 
fields $n_k$ may support solitons, provided 
the leading term (quadratic in derivatives $\pl_\mu n_k$) 
of the gradient expansion does not vanish \cite{sun1}.
The large N expansion seems a natural analytical method 
to verify this conjecture.

\subsubsection*{Acknowledgments}

I am grateful to L.D. Faddeev and J.R. Klauder for stimulating
and fruitful discussions. I thank the referee for drawing
my attention to works \cite{knot}.


\begin{thebibliography}\\  

\bibitem{faddeev1}L.D. Faddeev, {\em Quantization of Solitons},
IAS preprint, IAS-75-QS70, Princeton, 1970.
\bibitem{cho} Y.M. Cho, Phys. Rev. D 21 (1980) 1080; 23 (1981) 2415. 
\bibitem{faddeev2}L.D. Faddeev and A.J. Niemi, Phys. Rev. Lett.
82 (1999) 1624
\bibitem{f}L.D. Faddeev, {\em From Yang-Mills fields to solitons
and back again}, hep-th/9901037.
\bibitem{fn2}L.D. Faddeev and A.J. Niemi, Nature 378 (1997) 58.
\bibitem{knot} J. Gladikowski and M. Hellmund, Phys. Rev.
D56 (1997) 5194;\\
R.A. Battye and P.M. Sutcliffe, Phys. Rev. Lett. 81 (1998) 4798;\\
J. Hietarinta and P. Salo, Phys. Lett. B451 (1999) 60.
\bibitem{thooft}G. 't Hooft, Nucl. Phys. B190 [FS3] (1981) 455.
\bibitem{mag} A.S. Kronfeld, G. Schierholz and U.J. Wiese,
Nucl. Phys. B293 (1987) 461.
\bibitem{cam}Various aspects of Abelian projections in lattice
gauge theories are discussed by\\
M.N. Chernodub and M.I. Polikarpov, in: {\em Confinement,
Duality, and Nonperturbative aspects of QCD} (P. van Baal, ed.),
NATO ASI Series, Series B: Physics Vol 368, Plenum Press, 
London, 1998; p.378,\\
A. Di Giacomo, ibid.; p. 415,\\
T. Suzuki et al, ibid.; p. 439.
\bibitem{ad}T. Suzuki and I. Yotsuyanagi, Phys. Rev. D42 (1990) 4257.
\bibitem{md}J.D. Stack, S.D. Neiman and R.J. Wensley, 
Phys. Rev. D50 (1994) 3399.
\bibitem{wu}C.N. Yang and T.T. Wu, in: {\em Properties of Matter
Under Unusual Condition}, (H. Mark and S. Fernbach, eds) 
Interscience, New York, 1969.
\bibitem{im}O. Miyamura and S. Origuichi, in: {\em RCNP Confinement 1995},
Japan, Osaka, 1995; p.137.\\
M.N. Chernodub and F.V. Gubarev, JETP Lett. 62 (1995) 100;\\
A. Hart and M. Tepper, Phys. Lett. B371 (1996) 261;\\
M. Feuerstein, H. Markum and S. Thurner, Phys. Lett. B396 (1997) 203.
\bibitem{witten}E. Witten, Phys. Rev. Lett. 38 (1977) 121.
\bibitem{imc}R.H. Swendsen, Phys. Rev. Lett. 52 (1984) 1165;
Phys. Rev. D 30 (1984) 3866, 3875.
\bibitem{imcj}H. Shiba and T. Suzuki, Phys. Lett. B343 (1995) 315;
B351 (1995) 519.
\bibitem{sun1}L.D. Faddeev and A.J. Niemi, Phys. Lett. B449 (1999) 214.
\bibitem{sun2}V. Periwal, {\em Monopole condensates
in SU(N) Yang-Mills theory}, hep-th/9808127.
\bibitem{zhel} D.P. Zhelobenko, {\em Compact Lie Groups and Their
Representations}, Translations of mathematical monographs, Vol. 40,
AMS, Providence, RI,  1973.
\bibitem{svs} S.V. Shabanov, Mod. Phys. Lett. A 11 (1996) 1081;\\
{\em The monopole dominance in QCD}, hep-th/9611228.

\end{thebibliography}
\end{document}